\documentclass[twocolumn,showpacs,preprintnumbers,amsmath,amssymb,showkeys,aps,prb,reprint]{revtex4-1}
\usepackage{times}
\usepackage{epsfig}
\usepackage{hyperref}
\usepackage{bm}
\begin{document}

\title{Mechanism for the $\alpha \to \epsilon$ phase transition in iron}

\author{Bertrand Dup\'e,$^\dagger$ Bernard Amadon,$^{*}$ Yves-Patrick Pellegrini, and Christophe Denoual}
\affiliation{CEA, DAM, DIF, F-91297 Arpajon, France}

\date{Received 27 July 2012; published 7 January 2013}
\begin{abstract}
The mechanism of the $\alpha$-$\epsilon$ transition in iron is reconsidered. A path in the Burgers description of the bcc/hcp transition different from those previously considered is proposed. It relies on the assumption that shear and shuffle are decoupled and requires some peculiar magnetic order, different from that of $\alpha$ and $\epsilon$ phases as found in Density-Functional Theory. Finally, we put forward an original mechanism for this transition, based on successive shuffle motion of layers, which is akin to a nucleation-propagation process rather than to some uniform motion.
\end{abstract}

\pacs{64.70.kd, 75.50.Bb, 81.30.Kf
\hfill
\doi{10.1103/PhysRevB.87.024103}
}

\maketitle

\section{Introduction}

The bcc-hcp transition in iron has recently been the subject of intense experimental\cite{Wang98,Hawreliak06,Mathon2004,monza11} and theoretical\cite{Ekman98,Friak08,Johnson08,Lizarraga08,Liu2009,Okatov09,Toledano10,Leonov11,Sanchez09} works.
At room temperature and pressure, $\alpha$-iron is a bcc metal with ferromagnetic (FM) order. Upon pressure, iron exhibits a phase transition at $\simeq$13 GPa to hcp structure\cite{Takahashi1964,Clendenen1964,mao-272} with no magnetic order.\cite{Cort82} The hysteresis at the transition is large,\cite{Giles71} and as pressure is changed, the transformation occurs rapidly, suggesting that the transition is non-diffusive and martensitic. Recent work underlines the importance of antiferromagnetic (AFM) fluctuations in hcp $\epsilon$ iron \cite{monza11}.

In order to describe $\alpha$ iron from an ab-initio point of view, Dynamical Mean Field Theory\cite{Katanin10,Leonov11} in combination with Density Functional Theory (DFT) or other advanced schemes\cite{Sanchez09} have been used. In particular, the localization of electrons in the non-bonding e$_{\rm g}$ state but not in the bonding t$_{\rm 2g}$ state is a key point\cite{Katanin10} for understanding paramagnetic bcc iron. However, DFT ab initio techniques alone can handle FM bcc iron and the thermodynamical fundamentals of the $\alpha \to \epsilon$ transition\cite{Stixrude94}: The Generalized Gradient Approximation (GGA) correctly captures the basic Stoner mechanism for the reduction of magnetism, under the increase of atom coordination during the $\alpha \to \epsilon$ transition, as well as the larger spin polarization of e$_{\rm g}$ orbitals with respect to t$_{\rm 2g}$ in the $\alpha$ phase.

The $\alpha\to\epsilon$ transition can be viewed\cite{burgers1934} as combining an anisotropic compression (shear) in the (100) direction of bcc with a shuffle in the (011) plane, that corresponds to a zone-boundary transverse phonon mode (see, e.g, Ref. \onlinecite{Ekman98}). Due its martensitic nature, the transformation is triggered above the thermodynamical threshold, i.e., when the $\alpha$ and $\epsilon$ enthalpies are equal, which appears above $\simeq$ 10-13GPa,\cite{Ekman98,Johnson08,Liu2009} as computed in the GGA.

However, its ab-initio description is debated. Although the broad outline of the transition has been settled, a detailed description faces the problem of accurately computing the relevant energy barrier.
Ekman {\it et al.}\cite{Ekman98} demonstrated that the transition is first-order, with no dynamical precursors, and that it is caused by the effect of pressure on the magnetism of iron. Johnson and Carter\cite{Johnson08} have shown that a minimization wrt.\ shuffle and shear, considered as independent variables, leads to a cusp in the energy profile with a particularly low energy barrier. This pathway imposes a discontinuity of the shuffle displacement, at no additional energetic cost. Rejecting this possibility, Liu and Johnson \cite{Liu2009} stated that shuffle and shear are coupled and cannot be minimized separately. Still, the barrier energy found this way is too high ---within the range of pressures for which the transition is expected to occur--- to permit the transformation.\cite{Friak08} No mechanism allowing for a transformation with a reasonable energy barrier has been proposed.

Investigations have established the role of complex magnetic structures for some transition paths of the $\alpha \to \gamma$ (Refs.\ \onlinecite{Okatov09}, \onlinecite{Tsetseris05} and \onlinecite{Tsunoda89}) and $\alpha \to \epsilon$ transitions,\cite{Friak08} e.g., for $\epsilon-$Fe spin spiral states have the same energy as AFM structures.\cite{Lizarraga08} Yet no study has been carried out on the effects of magnetic ordering on the pathway of the bcc-hcp martensitic phase transition.

As to the kinetics of the transition, no intermediate states were detected in experiments on the nanosecond time scale,\cite{Hawreliak11} which suggests that the transformation is very rapid and probably propagates at nearly sound velocity. Moreover, it has been shown recently that for cobalt, the transition occurs through microstructural avalanches.\cite{Sanborn11} Such behavior implies that the transition is not of the military type but instead involves localized transformation events. No such possibility has been considered yet for iron.

In this paper, we first propose (Sec.\ \ref{sec:A}) an alternative description of the path for the $\alpha \to \epsilon$ transition within the Burgers mechanism. Second, using GGA calculations, we show the importance of AFM order to describe the energetics of this path (Sec.\ \ref{sec:B1}). Third, and this is our main result, we emphasize (Secs.\ \ref{sec:B2} and \ref{sec:C}) that the shuffle mechanism could occur layer after layer in a non-simultaneous way, thus, bridging the gap between the low-energy pathway of Ref. \onlinecite{Johnson08} and a description of the transition without any discontinuity in displacement.\cite{Liu2009}

\section{Computational Setup and Methodology}
\label{sec:A}
\begin{figure}
\includegraphics[width=5.0cm]{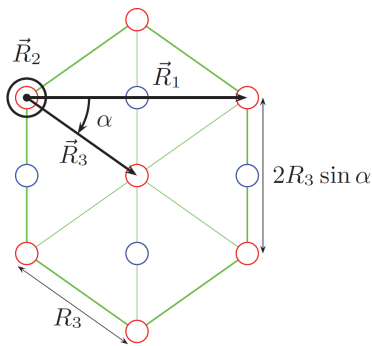}
\caption{(Color online) Scheme of the $(001)_{\text{hcp}}\parallel (110)_{\text{bbc}}$ plane in the bcc configuration. The vectors $\vec{R}_1$, $\vec{R}_2$, and $\vec{R}_3$ define our simulation cell. The red atoms and blue atoms lie in two different planes (see Fig. \ref{fig:path}). $\vec{R}_2$ is perpendicular to $\vec{R}_1$ and $\vec{R}_3$, and $R_2=4c$. The shear is defined by $\epsilon=(2\sin{\alpha}-\frac{2}{\sqrt{3}})/(1-\frac{2}{\sqrt{3}})$. In bcc, $2\sin{\alpha}=\frac{2}{\sqrt{3}}$, and $\epsilon=0$. In hcp, $2\sin{\alpha}=1$, thus, $\epsilon=1$. From this definition, conservation of volume ($R_2R_1R_3\sin{\alpha}=V$), conservation of $\frac{c}{a}$ ($=\frac{R_2}{4R_3}=\sqrt{\frac{8}{3}}$), and the geometric relation $2R_3\cos{\alpha}=R_1$, give $\vec{R}_1$, $\vec{R}_2$, and $\vec{R}_3$ as a function of the value of the shear.}
\label{fig:cell}.
\end{figure}

\begin{figure}
\includegraphics[width=8.5cm]{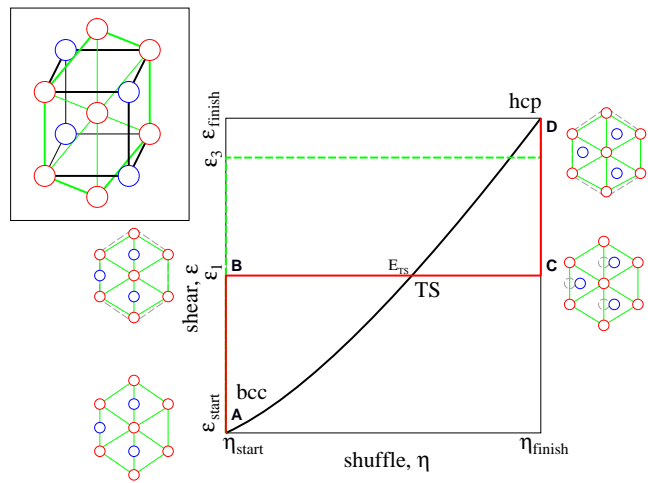}
\caption{(Color online) Sketch of possible paths for the bcc/hcp transition in the shear shuffle plane.
TS is the transition state, and the black line corresponds to the lowest pass to go from the bcc valley to the hcp valley. The red and green lines are the transition paths that are studied in Fig. \ref{fig:energyshuffle} (see also text). $\epsilon_1=0.5$ and $\epsilon_3=0.875$.}
\label{fig:path}
\end{figure}

The two-atom unit cell is replicated four times in the $(001)_{\text{hcp}}\parallel (110)_{\text{bbc}}$ direction into an eight-atom supercell, with periodic boundary conditions (PBCs). The unit cell is represented in Fig. \ref{fig:cell} with lattice vectors $\vec{R}_1$, $\vec{R}_2$, and $\vec{R}_3$. Three parameters are necessary to determine the unit cell: the angle $\alpha$, the value of $c=R_2/4$ and the value of $R_1$.
Three macroscopic quantities can be used to compute them: the volume, the $c/a=R_2/(4R_3)$ ratio and the shear $\epsilon$. In our paper, the volume, namely, 71.5 bohr${}^3$/atom, is such that hcp and bcc energies coincide. The $c/a$ ratio is kept constant and equal to $\sqrt{8/3}\simeq1.633$, which is the value in the bcc structure and in the ideal hcp compact structure. The shear is defined  by $\epsilon=(2\sin{\alpha}-\frac{2}{\sqrt{3}})/(1-\frac{2}{\sqrt{3}})$ and
is equal respectively to 0 and 1, respectively, in the bcc and the hcp structures.
For each value of the shear, we computed the energy for three values of the $c/a$ ratio, namely, 1.605, $\sqrt{8/3}\simeq 1.633$, and 1.668 to check that the sheared bcc structure is stable with respect to a change of $c/a$ and that $c/a=\sqrt{8/3}$ is close to the minimum. The results are insensitive to the small difference between this value and the true minimum. Finally, the shuffle $\eta$ is proportional to the distance of one blue atom between its position during the transition (see Fig.\ref{fig:path}) and the reference position in bcc.

We focus our paper on the shuffle part of the Burgers mechanism. Because of the different physical timescales involved,\cite{Johnson08} we assume that the lattice shear deformation mode ($\epsilon$) and the shuffle mode ($\eta$) are uncoupled in such manner that, with respect to shear, shuffle is instantaneous. Volume and $c/a$ ratio are, thus, kept constant during the shuffle. The path proposed in Ref.\ \onlinecite{Johnson08} is defined by the equality of the energies of the sheared bcc and hcp phases. In Ref.\ \onlinecite{Johnson08}, however, the energetics of the shuffle is not explored. On the other hand, Liu and Johnson\cite{Liu2009}
couple shear and shuffle modes. Their path is schematically represented in black in Fig.\ref{fig:path}.
In this figure, the shear ($\epsilon_1$) is fixed to half the value necessary to go from the bcc $\alpha$ phase to the hcp $\epsilon$ phase. It is the shear at the transition state computed by Ekman et al.\cite{Ekman98} at constant volume.  This choice guarantees that the shuffle mode goes through the transition state (TS) of Ref. \onlinecite{Ekman98} and  corresponds to the energetically most favorable path, represented in solid red in Fig.\ref{fig:path}.

In this paper, we are concerned by the shuffle mechanism along the BC line (see letters in Fig.\ \ref{fig:path}). Since the shuffle of a large number of atoms can be described in several different ways, our aim is to understand whether the shuffle of all the layers is simultaneous or not.

\subsection{Simultaneous shuffle}
\begin{figure}
\includegraphics[width=8.5cm]{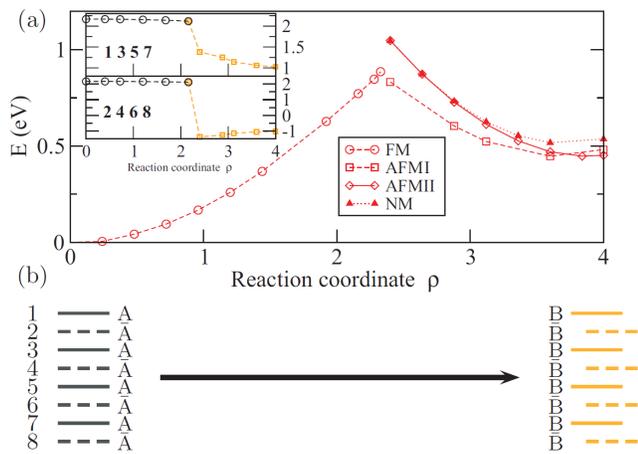}
\caption{(Color online) (a) Total energy for the simultaneous shuffle $S_1$ computed for
FM, AFMI, AFMII,\cite{Note2} and Non-Magnetic (NM) configurations. $S$ stands for Simultaneous and the index 1 refers to the shear $\epsilon_1$ (see Fig.\ \ref{fig:path}). The energy is given for a supercell of eight atoms to allow for an easy comparison with Fig.\ \ref{fig:energyshuffle}. (b) Stacking of the layers during the transition. In black, the A and $\bar{\rm A}$ layers. In orange, the B and $\bar{\rm B}$ layers. The black (respectively, orange) color, thus, indicates that the first shell of neighbors is bcc- (respectively, hcp-) like. In the inset, the evolution of the magnetic moment for the eight atoms during the FM $\rightarrow$ AFMI transition.}
\label{fig:afmii}
\end{figure}

Let us first consider simultaneous shuffle (S). By ${\rm A\bar{\rm A}A\bar{\rm A}A\bar{\rm A}A\bar{\rm A}}$, we denote the stacking of the eight atoms in the bcc sheared phase, and by ${\rm B\bar{\rm B}B\bar{\rm B}B\bar{\rm B}B\bar{\rm B}}$, we denote their stacking in the hcp sheared phase. In this notation, A and $\bar{\rm A}$, and B and $\bar{\rm B}$, correspond to atoms in the bcc- or hcp-like configurations, respectively. The atoms in the ${\rm A}$ and $\bar{\rm A}$  layers have 8 neighbors, whereas atoms in the ${\rm B}$ and $\bar{\rm B}$ layer have 12 neighbors. During the simultaneous shuffling of all four atoms of the $\bar{\rm A}$-type layer into  a $\bar{\rm B}$-type layer, A-type layers are transformed into B-type layers. The transformation is, thus, ${\rm A\bar{\rm A}A\bar{\rm A}A\bar{\rm A}A\bar{\rm A}}$ $\rightarrow$  ${\rm B\bar{\rm B}B\bar{\rm B}B\bar{\rm B}B\bar{\rm B}}$ [Fig. \ref{fig:afmii}(b)].

\subsection{Consecutive shuffle}
Consider next the consecutive individual shuffle of layers (C), schematically represented in Fig.\ \ref{fig:energyshuffle}(b). First, we compute the individual shuffling of layer 2 from $\bar{\rm A}$ to $\bar{\rm B}$. It corresponds to the transformation ${\rm A\bar{\rm A}A\bar{\rm A}A\bar{\rm A}A\bar{\rm A}}$ $\rightarrow$ ${\rm I\bar{\rm B}I\bar{\rm A}A\bar{\rm A}A\bar{\rm A}}$. Due to the PBCs, this creates
intermediate (I) configurations for the first and third layers. It corresponds to an atom layer sandwiched between one ${\bar{\rm B}}$ and one ${\bar{\rm A}}$ layer. Their local environment is, thus, neither bcc nor hcp.

In a second step, starting from the latter configuration, we study the shuffling of the second layer up to the new configuration ${\rm I\bar{\rm B}B\bar{\rm B}I\bar{\rm A}A\bar{\rm A}}$.  During this transformation, the fourth layer shuffles from configuration $\bar{\rm A}$ to configuration $\bar{\rm B}$. Thus, the third layer now has a full hcp-like configuration around it and now is, thus, labeled B. The process can be pursued until all layers have shuffled.

\subsection{Computational details}
We use the Projector Wave Augmented\cite{PhysRevB.50.17953} implementation\cite{Torrent2008337}
of {\sc ABINIT} (Ref.\ \onlinecite{Gonze09}) in the GGA PBE approximation of DFT. Atomic data include 3s and 3p semi-core states,\cite{PhysRevB.50.17953} the cutoff radius is 2.0 a.u., the energy cutoff for the plane-wave expansion is 20 Ha, and the convergence criterion for the charge-density residual is 10$^{-9}$. With these atomic data, results of Ref.\ \onlinecite{Torrent2008337} are reproduced. A 18x4x26 special $k$-point mesh\cite{PhysRevB.13.5188} and a Gaussian smearing of $5.10^{-4}$ Ha for electronic occupations\cite{PhysRevB.40.3616} are used in order to obtain a good estimate of the energetic and magnetic properties.

\section{Results}
\subsection{Simultaneous shuffle}
\label{sec:B1}
Energies for the simultaneous mechanism are displayed in Fig.\ \ref{fig:afmii}(a). The reaction coordinate ($\rho$) is taken proportional to the shuffle ($\eta$) of the supercell atoms and varies from 0 (bcc-like) to 4 (all atoms in hcp-like environment). Near equilibrium ($\rho=0$), the  energy increases quadratically as expected. The curve exhibits a cusp at 60\% shuffle amplitude due to a change of magnetism from FM to AFMI.\cite{Note2} During this transition, the magnetic moment---reproduced in the inset of Fig.\ \ref{fig:afmii}---changes abruptly from $2.2 \mu_B/{\rm atom}.$ to $1.1 \mu_B/{\rm atom}$ in agreement with Ref.\ \onlinecite{Friak08} where, however, AFMI order was not considered. This confirms the importance of magnetism for this transition:\cite{Ekman98,Friak08,Liu2009,Johnson08} Without any change in magnetism, the transition would not occur. In the rightmost part of the curve, the energy decreases to a local minimum corresponding to a deformed hcp phase. We computed the same curve with a cell of four atoms allowing for the creation of AFMII order.\cite{Note2} We find, in agreement with Refs.\ \onlinecite{Steinle99} and \onlinecite{Lizarraga08}, that AFMII is more stable for the hcp structure ($\rho$=4) compared to AFMI. However, and interestingly, we also find that past the cusp, the AFMI structure is more stable than the AFMII. This highlights the important role of complex magnetism for the DFT description of the transition path and not only for the stable structure.\cite{Ekman98} A complete study of magnetism, especially noncollinear, lies beyond the goal of our paper. We nevertheless show that an AFMI description of the transition path is particularly adapted to the present case. We call $E^{\rm S_1}({\rho})$ the combination of the FM and AFMI curves. The energy profile qualitatively agrees with previous results, involving a smaller unit cell with two atomic layers (Figs.\ 5 and 6 of Ref. \onlinecite{Ekman98}) even if Ekman {\it et al.} have considered a NM ground state for the hcp phase and the PW91 GGA functional.

\subsection{Consecutive shuffle}
\label{sec:B2}
\begin{figure}
\includegraphics[width=8.5cm]{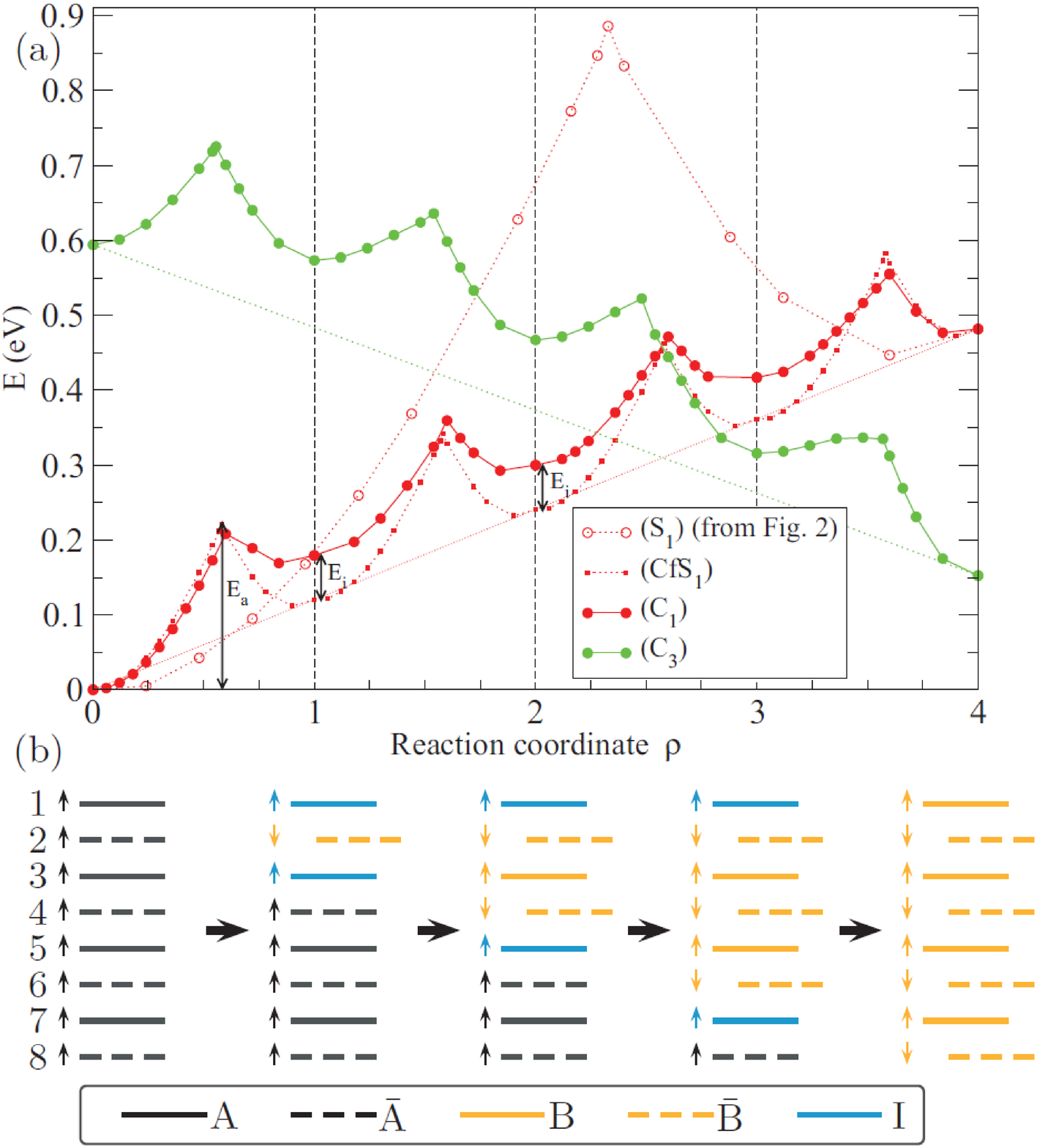}
\caption{(Color online) (a) Comparison of energies for three different shuffling mechanisms (for a shear of $\epsilon_1$=0.5). (S$_1$) is the simultaneous shuffling of all atoms. (CfS$_1$) stands for ``consecutive from simultaneous",  an hypothetical mechanism that corresponds to consecutive shuffles within an assumption of independent layers (interactions neglected), computed with the energies of mechanism $S_1$ (see text). (C$_1$) is the true energy computed when layers shuffle in a consecutive way (interactions included). $E_i$ and $E_a$ represent, respectively, the interface energy and the activation energy for the first shuffle (see text).
(C$_3$) is the energy for the consecutive shuffling of the atoms for a shear of $\epsilon_3$=0.875.
(b) Layer stacking during consecutive shuffling: in blue, layers in which the first atomic shell is intermediate between the bcc and hcp environments. Arrows indicate the sign of the magnetic moment in each layer, as described in Fig.\ \ref{fig:localmag}.}
\label{fig:energyshuffle}
\end{figure}

Energy plots associated with the consecutive mechanism are drawn in Fig.\ \ref{fig:energyshuffle}(a). From $E^{\rm S_1}({\rho})$, one can trivially compute the energy of shuffle per atom and, thus, can simulate the consecutive motion of four layers under the assumption that the energy for the shuffling of only one layer
is equal to that for shuffling all layers, divided by the number of layers. This approximation would be correct if there were no interactions between layers. For this mechanism, because all atoms
are not shuffling at the same time, $\rho$ describes the consecutive shuffling of all the layers.
Formally, from value $n-1$ to value $n$, $({\rho}-(n-1))$ is, thus, proportional to the displacement
of the atoms of the $n^{\rm th}$ layer. As a consequence, for $\rho$=0 and $\rho$=4, whatever the mechanism, all atoms are in the same positions. The expression for $E^{\rm CfS}({\rho})$ between $n-1$ and $n$ is as follows:
\begin{equation}
\nonumber
E^{\rm CfS}({\rho})=\frac{E^{\rm S}(4({\rho}-(n-1)))}{4}+(n-1)\frac{E^{\rm S}({\rho}=4)}{4}.
\end{equation}
As a consequence, $E^{\rm CfS}(0)$, $E^{\rm CfS}(1)$, $E^{\rm CfS}(2)$, $E^{\rm CfS}(3)$,
and $E^{\rm CfS}(4)$ lie on the same line (dashed red in Fig.\ \ref{fig:energyshuffle}). Trivially, this mechanism reduces the total energy barrier: In Fig. \ref{fig:energyshuffle}, the energy barrier
$E^{\rm CfS_1}(\rho\simeq3.53)- E^{\rm CfS_1}(\rho=0)$ is lower than the energy barrier
$E^{\rm S_1}(\rho\simeq2.30)-E^{\rm S_1}(\rho=0)$. Finally, the third curve ($E^{\rm C_1}$) corresponds to
the explicit calculation of the consecutive motion of four layers.

\subsection{Discussion}
\label{sec:C}
We focus next on the comparison between the model of independent layers (CfS$_1$) and the true calculation (C$_1$). The two curves show the same main tendencies with four energy barriers corresponding to the shuffling of the four atoms and three metastable structures corresponding to the shuffling of only one, two or three layers (${\rho}=1$--$3$). However, the energy of the three metastable configurations (for ${\rho}=1$--$3$) differs by only $E_i=E^{\rm C_1}(1)-E^{\rm CfS_1}(1)$ = 60 meV and is, thus, a constant independent of the number of shuffled layers. This suggests that $E_i$ comes from the two interfaces (in blue in Fig.\ \ref{fig:energyshuffle}) between shuffled and unshuffled layers. The number of such interfaces is constant and independent of the number of layers.

\begin{figure}
\includegraphics[width=8.5cm]{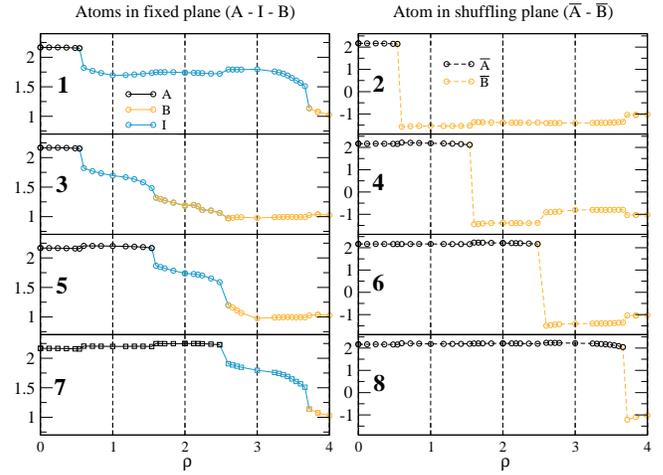}
\caption{(Color online) Evolution of local magnetic moment for the eight atoms of the supercell as a function of the reaction coordinate $\rho$. The black, orange, and blue colors approximately depict the bcc, hcp and mixed bcc/hcp values for the magnetic moment. The atom numbering is related to Fig.\ \ref{fig:energyshuffle}. Atoms 1, 3, 5, 7 belong to the A layers and atoms 2, 4, 6, 8 are successively moving from and $\bar{\rm A}$-type layer to $\bar{\rm B}$-type one.}
\label{fig:localmag}
\end{figure}

To assess the electronic origin of these interface energies, Fig.\ \ref{fig:localmag} represents
the evolution of the d-magnetic moment for the eight layers in the supercell as a function of $\rho$. The transition goes along with a progressive change of magnetism in which each successive layer shuffle, from bcc- to hcp-like configuration, involves local jumps from FM to AFMI. Indeed, after plane number 2 is translated,
there is a local AFMI order around this layer. When plane number 4 is translated, AFMI order expands over 5 layers, and so forth. Also, the local moments strongly depend on local coordination, in agreement with
the generalized Stoner mechanism for the appearance of magnetism as a function of bandwidth.\cite{Ekman98}
For example, when the fourth atom goes from a bcc-like local configuration to a more compact hcp-like local configuration, its local moment changes from 2.2 $\mu_B$ to -1.1 $\mu_B$ with no substantial difference with the simultaneous mechanism. At the same time, the moment of the neighboring atom 5 changes from 2.2 $\mu_B$
to 1.7 $\mu_B$ only, because atom 5 has now a local environment intermediate between the hcp and the bcc.
Once atom 6 has moved, atom 5 ends up (for $\rho$=3) with a full hcp-like local environment and a moment of 1.1 $\mu_B$. This demonstrates: (i) the absolute correlation between magnetism and total energy in the transition and (ii) the link between interfacial energy and layers with intermediate local moment and coordination. The interfacial energy only depends on the number of interfaces, which is constant in the present paper. Remarkably, energy barriers are not shifted by this interfacial energy.

For $\epsilon_1=0.5$, one notices that the energy progressively increases for all the metastable states up to the final energy gain of -0.5eV. An additional calculation is performed for a shear of $\epsilon_3 = 0.875$ (green lines in Fig.\ \ref{fig:path} and Fig.\ \ref{fig:energyshuffle}). The resulting energy exhibits the same characteristics as for $\epsilon_1=0.5$, and notably an interface energy of 100 meV but with an
overall gain after shuffle of +0.45 eV. For $\epsilon_3 = 0.875$ and only around $\rho=3$, a new magnetic order---not stable for $\epsilon_1 = 0.5$---diminishes the interface energy to 52 meV. It embodies both FM and AFM interactions between the layers and highlights the energetic proximity of several magnetic orders. Further studies, with larger supercells, would be necessary to study this magnetic order, but this is not the goal
of the present paper.

Two shear thresholds for the bulk transformation can be estimated. A simple linear interpolation between the shuffle curves C$_1$ and C$_3$ produces the following energy for a shear $\epsilon_2$ such that $\epsilon_1 \leq \epsilon_2 \leq \epsilon_3$ and the shuffle $\rho$:
\begin{equation}
\nonumber
E(\rho,\epsilon_2)=E^{\rm C_1}(\rho)+
\frac{E^{\rm C_3}(\rho)-E^{\rm C_1}(\rho)}
{\epsilon_3-\epsilon_1}
(\epsilon_2-\epsilon_1).
\end{equation}
A lower nucleation-threshold (LT) shear $\epsilon_2^{\rm LT}\simeq0.7$ is defined as the shear above which the gain $E(\rho=0,\epsilon_2)-E(\rho=4,\epsilon_2)$ becomes positive. To complete the transformation, the interface energy has to be overcome for the first layer only, leading to shuffling of subsequent layers at no additional cost. For a large number of layers, thermodynamical equilibrium, thus, drives the transition to completion. Also, an upper-nucleation threshold (UT) shear of $\epsilon_2^{\rm UT}\simeq0.84$ is estimated from the linear interpolation. It is defined as the shear above which the gain $E(\rho=0,\epsilon_2)-E(\rho=1,\epsilon_2)$ for the first metastable state becomes positive.
Above this shear, even the first shuffle occurs at no energy cost. Then, all subsequent shuffles are strongly favored, possibly leading to propagation and to a global transition.

Consequences on the kinetics of the $\alpha$-$\epsilon$ transition are as follows. Shuffle can be viewed as a thermally-activated kinetic process due to vibrations of atoms.\cite{Ekman98,Johnson08,Liu2009}
As emphasized above, the energy barrier per atom $E_a$, which defines the activation energy, is independent of the interfacial energy so that the shuffle of a single layer has, thus, no supplementary activation cost.
In this perspective, the above provides an alternative mechanism for the transition in which the bulk shuffle transformation nucleates from one single shuffle event, possibly initiated by some initial shear $\epsilon_2$, such as, e.g., in a shock wave. This mechanism considerably reduces the amount of energy needed to trigger the transformation, the dimensionality of the nucleation process being reduced from 3 to 2.

In order to describe a transition corresponding to a shuffle for a domain, a complete thermodynamical description of this mechanism would be necessary. This would require relaxing atoms after each shuffle, and is beyond the scope of our paper.

\section{Conclusion}
In conclusion, we utilize the fact that shuffle---or optical phonon mechanism---and macroscopic compression
---shear---have in general different timescales, which allows for their decoupling. This is true for iron but also for other systems such as Zr, Ba, or related transitions, such as Pu.\cite{Lookman2008} Second, in iron bcc-hcp transition, we show that the transition-path description requires some peculiar magnetic order different from the magnetic order of equilibrium phases. Third, we highlight a new mechanism for this transition based on  successive shuffle motion of layers. It indicates that each layer itself could move by developing ``shuffle dislocations" in a manner akin to slip motion in plasticity. Our paper suggests that detailed investigations of nucleation mechanisms would be required to resolve the frontier between the military and the thermally activated characters of solid-solid phase transitions.

\acknowledgments
We thank Marc Torrent for useful discussions and Agn\`es Dewaele for useful discussions and a careful reading of the paper. This work was granted access to the HPC resources of CCRT and CINES under the Allocations No.\ 2011096681 and No.\ 2012096681 made by GENCI (Grand \'Equipement National de Calcul Intensif).\\

{\small $\dagger$ Present adress: Institut f\"ur Theoretische Physik und Astrophysik, Christian-Albrechts-Universit\"at zu Kiel, Leibnizstrasse 15, D-24098 Kiel, Germany.}\\
{\small $^{*}$ bernard.amadon@cea.fr}

\begin{thebibliography}{34}
\expandafter\ifx\csname natexlab\endcsname\relax\def\natexlab#1{#1}\fi
\expandafter\ifx\csname bibnamefont\endcsname\relax
  \def\bibnamefont#1{#1}\fi
\expandafter\ifx\csname bibfnamefont\endcsname\relax
  \def\bibfnamefont#1{#1}\fi
\expandafter\ifx\csname citenamefont\endcsname\relax
  \def\citenamefont#1{#1}\fi
\expandafter\ifx\csname url\endcsname\relax
  \def\url#1{\texttt{#1}}\fi
\expandafter\ifx\csname urlprefix\endcsname\relax\def\urlprefix{URL }\fi
\providecommand{\bibinfo}[2]{#2}
\providecommand{\eprint}[2][]{\url{#2}}

\bibitem[{\citenamefont{Wang and Ingalls}(1998)}]{Wang98}
\bibinfo{author}{\bibfnamefont{F.~M.} \bibnamefont{Wang}} \bibnamefont{and}
  \bibinfo{author}{\bibfnamefont{R.}~\bibnamefont{Ingalls}},
  \bibinfo{journal}{Phys. Rev. B} \textbf{\bibinfo{volume}{57}},
  \bibinfo{pages}{5647} (\bibinfo{year}{1998}),

\bibitem[{\citenamefont{Hawreliak et~al.}(2006)\citenamefont{Hawreliak, Colvin,
  Eggert, Kalantar, Lorenzana, S.~St\"olken, Davies, Germann, Holian, Kadau
  et~al.}}]{Hawreliak06}
\bibinfo{author}{\bibfnamefont{J.}~\bibnamefont{Hawreliak}},
  \bibinfo{author}{\bibfnamefont{J.~D.} \bibnamefont{Colvin}},
  \bibinfo{author}{\bibfnamefont{J.~H.} \bibnamefont{Eggert}},
  \bibinfo{author}{\bibfnamefont{D.~H.} \bibnamefont{Kalantar}},
  \bibinfo{author}{\bibfnamefont{H.~E.} \bibnamefont{Lorenzana}},
  \bibinfo{author}{\bibfnamefont{J.}~\bibnamefont{S.~St\"olken}},
  \bibinfo{author}{\bibfnamefont{H.~M.} \bibnamefont{Davies}},
  \bibinfo{author}{\bibfnamefont{T.~C.} \bibnamefont{Germann}},
  \bibinfo{author}{\bibfnamefont{B.~L.} \bibnamefont{Holian}},
  \bibinfo{author}{\bibfnamefont{K.}~\bibnamefont{Kadau}},
  \bibnamefont{et~al.}, \bibinfo{journal}{Phys. Rev. B}
  \textbf{\bibinfo{volume}{74}}, \bibinfo{pages}{184107}
  (\bibinfo{year}{2006}).

\bibitem[{\citenamefont{Mathon et~al.}(2004)\citenamefont{Mathon, Baudelet,
  Iti\'e, Polian, d'Astuto, Chervin, and Pascarelli}}]{Mathon2004}
\bibinfo{author}{\bibfnamefont{O.}~\bibnamefont{Mathon}},
  \bibinfo{author}{\bibfnamefont{F.}~\bibnamefont{Baudelet}},
  \bibinfo{author}{\bibfnamefont{J.~P.} \bibnamefont{Iti\'e}},
  \bibinfo{author}{\bibfnamefont{A.}~\bibnamefont{Polian}},
  \bibinfo{author}{\bibfnamefont{M.}~\bibnamefont{d'Astuto}},
  \bibinfo{author}{\bibfnamefont{J.~C.} \bibnamefont{Chervin}},
  \bibnamefont{and}
  \bibinfo{author}{\bibfnamefont{S.}~\bibnamefont{Pascarelli}},
  \bibinfo{journal}{Phys. Rev. Lett.} \textbf{\bibinfo{volume}{93}},
  \bibinfo{pages}{255503} (\bibinfo{year}{2004}).

\bibitem[{\citenamefont{Monza et~al.}(2011)\citenamefont{Monza, Meffre,
  Baudelet, Rueff, d'Astuto, Munsch, Huotari, Lachaize, Chaudret, and
  Shukla}}]{monza11}
\bibinfo{author}{\bibfnamefont{A.}~\bibnamefont{Monza}},
  \bibinfo{author}{\bibfnamefont{A.}~\bibnamefont{Meffre}},
  \bibinfo{author}{\bibfnamefont{F.}~\bibnamefont{Baudelet}},
  \bibinfo{author}{\bibfnamefont{J.-P.} \bibnamefont{Rueff}},
  \bibinfo{author}{\bibfnamefont{M.}~\bibnamefont{d'Astuto}},
  \bibinfo{author}{\bibfnamefont{P.}~\bibnamefont{Munsch}},
  \bibinfo{author}{\bibfnamefont{S.}~\bibnamefont{Huotari}},
  \bibinfo{author}{\bibfnamefont{S.}~\bibnamefont{Lachaize}},
  \bibinfo{author}{\bibfnamefont{B.}~\bibnamefont{Chaudret}}, \bibnamefont{and}
  \bibinfo{author}{\bibfnamefont{A.}~\bibnamefont{Shukla}},
  \bibinfo{journal}{Phys. Rev. Lett.} \textbf{\bibinfo{volume}{106}},
  \bibinfo{pages}{247201} (\bibinfo{year}{2011}).

\bibitem[{\citenamefont{Ekman et~al.}(1998)\citenamefont{Ekman, Sadigh,
  Einarsdotter, and Blaha}}]{Ekman98}
\bibinfo{author}{\bibfnamefont{M.}~\bibnamefont{Ekman}},
  \bibinfo{author}{\bibfnamefont{B.}~\bibnamefont{Sadigh}},
  \bibinfo{author}{\bibfnamefont{K.}~\bibnamefont{Einarsdotter}},
  \bibnamefont{and} \bibinfo{author}{\bibfnamefont{P.}~\bibnamefont{Blaha}},
  \bibinfo{journal}{Phys. Rev. B} \textbf{\bibinfo{volume}{58}},
  \bibinfo{pages}{5296} (\bibinfo{year}{1998}).

\bibitem[{\citenamefont{Fri\'ak and $\check{S}$ob}(2008)}]{Friak08}
\bibinfo{author}{\bibfnamefont{M.}~\bibnamefont{Fri\'ak}} \bibnamefont{and}
  \bibinfo{author}{\bibfnamefont{M.}~\bibnamefont{$\check{S}$ob}},
  \bibinfo{journal}{Phys. Rev. B} \textbf{\bibinfo{volume}{77}},
  \bibinfo{pages}{174117} (\bibinfo{year}{2008}).

\bibitem[{\citenamefont{Johnson and Carter}(2008)}]{Johnson08}
\bibinfo{author}{\bibfnamefont{D.~F.} \bibnamefont{Johnson}} \bibnamefont{and}
  \bibinfo{author}{\bibfnamefont{E.~A.} \bibnamefont{Carter}},
  \bibinfo{journal}{J. Chem. Phys.}
  \textbf{\bibinfo{volume}{128}}, \bibinfo{eid}{104703}
  (\bibinfo{year}{2008}).

\bibitem[{\citenamefont{Liz\'arraga et~al.}(2008)\citenamefont{Liz\'arraga,
  Nordstr\"om, Eriksson, and Wills}}]{Lizarraga08}
\bibinfo{author}{\bibfnamefont{R.}~\bibnamefont{Liz\'arraga}},
  \bibinfo{author}{\bibfnamefont{L.}~\bibnamefont{Nordstr\"om}},
  \bibinfo{author}{\bibfnamefont{O.}~\bibnamefont{Eriksson}}, \bibnamefont{and}
  \bibinfo{author}{\bibfnamefont{J.}~\bibnamefont{Wills}},
  \bibinfo{journal}{Phys. Rev. B} \textbf{\bibinfo{volume}{78}},
  \bibinfo{pages}{064410} (\bibinfo{year}{2008}).

\bibitem[{\citenamefont{Liu and Johnson}(2009)}]{Liu2009}
\bibinfo{author}{\bibfnamefont{J.~B.} \bibnamefont{Liu}} \bibnamefont{and}
  \bibinfo{author}{\bibfnamefont{D.~D.} \bibnamefont{Johnson}},
  \bibinfo{journal}{Phys. Rev. B} \textbf{\bibinfo{volume}{79}},
  \bibinfo{pages}{134113} (\bibinfo{year}{2009}).

\bibitem[{\citenamefont{Okatov et~al.}(2009)\citenamefont{Okatov, Kuznetsov,
  Gornostyrev, Urtsev, and Katsnelson}}]{Okatov09}
\bibinfo{author}{\bibfnamefont{S.~V.} \bibnamefont{Okatov}},
  \bibinfo{author}{\bibfnamefont{A.~R.} \bibnamefont{Kuznetsov}},
  \bibinfo{author}{\bibfnamefont{Y.~N.} \bibnamefont{Gornostyrev}},
  \bibinfo{author}{\bibfnamefont{V.~N.} \bibnamefont{Urtsev}},
  \bibnamefont{and} \bibinfo{author}{\bibfnamefont{M.~I.}
  \bibnamefont{Katsnelson}}, \bibinfo{journal}{Phys. Rev. B}
  \textbf{\bibinfo{volume}{79}}, \bibinfo{pages}{094111}
  (\bibinfo{year}{2009}).

\bibitem[{\citenamefont{Tol\'edano et~al.}(2010)\citenamefont{Tol\'edano,
  Katzke, and Machon}}]{Toledano10}
\bibinfo{author}{\bibfnamefont{P.}~\bibnamefont{Tol\'edano}},
  \bibinfo{author}{\bibfnamefont{H.}~\bibnamefont{Katzke}}, \bibnamefont{and}
  \bibinfo{author}{\bibfnamefont{D.}~\bibnamefont{Machon}},
  \bibinfo{journal}{J. Phys.: Condens. Matter}
  \textbf{\bibinfo{volume}{22}}, \bibinfo{pages}{466002}
  (\bibinfo{year}{2010}).

\bibitem[{\citenamefont{Leonov et~al.}(2011)\citenamefont{Leonov, Poteryaev,
  Anisimov, and Vollhardt}}]{Leonov11}
\bibinfo{author}{\bibfnamefont{I.}~\bibnamefont{Leonov}},
  \bibinfo{author}{\bibfnamefont{A.~I.} \bibnamefont{Poteryaev}},
  \bibinfo{author}{\bibfnamefont{V.~I.} \bibnamefont{Anisimov}},
  \bibnamefont{and}
  \bibinfo{author}{\bibfnamefont{D.}~\bibnamefont{Vollhardt}},
  \bibinfo{journal}{Phys. Rev. Lett.} \textbf{\bibinfo{volume}{106}},
  \bibinfo{pages}{106405} (\bibinfo{year}{2011}).

\bibitem[{\citenamefont{S\'anchez-Barriga
  et~al.}(2009)\citenamefont{S\'anchez-Barriga, Fink, Boni, Di~Marco, Braun,
  Min\'ar, Varykhalov, Rader, Bellini, Manghi et~al.}}]{Sanchez09}
\bibinfo{author}{\bibfnamefont{J.}~\bibnamefont{S\'anchez-Barriga}},
  \bibinfo{author}{\bibfnamefont{J.}~\bibnamefont{Fink}},
  \bibinfo{author}{\bibfnamefont{V.}~\bibnamefont{Boni}},
  \bibinfo{author}{\bibfnamefont{I.}~\bibnamefont{Di~Marco}},
  \bibinfo{author}{\bibfnamefont{J.}~\bibnamefont{Braun}},
  \bibinfo{author}{\bibfnamefont{J.}~\bibnamefont{Min\'ar}},
  \bibinfo{author}{\bibfnamefont{A.}~\bibnamefont{Varykhalov}},
  \bibinfo{author}{\bibfnamefont{O.}~\bibnamefont{Rader}},
  \bibinfo{author}{\bibfnamefont{V.}~\bibnamefont{Bellini}},
  \bibinfo{author}{\bibfnamefont{F.}~\bibnamefont{Manghi}},
  \bibnamefont{et~al.}, \bibinfo{journal}{Phys. Rev. Lett.}
  \textbf{\bibinfo{volume}{103}}, \bibinfo{pages}{267203}
  (\bibinfo{year}{2009}).

\bibitem[{\citenamefont{Takahashi and Bassett}(1964)}]{Takahashi1964}
\bibinfo{author}{\bibfnamefont{T.}~\bibnamefont{Takahashi}} \bibnamefont{and}
  \bibinfo{author}{\bibfnamefont{W.~A.} \bibnamefont{Bassett}},
  \bibinfo{journal}{Science} \textbf{\bibinfo{volume}{145}},
  \bibinfo{pages}{483} (\bibinfo{year}{1964}).

\bibitem[{\citenamefont{Clendenen and Drickamer}(1964)}]{Clendenen1964}
\bibinfo{author}{\bibfnamefont{R.}~\bibnamefont{Clendenen}} \bibnamefont{and}
  \bibinfo{author}{\bibfnamefont{H.}~\bibnamefont{Drickamer}},
  \bibinfo{journal}{J. Phys. Chem. Solids}
  \textbf{\bibinfo{volume}{25}}, \bibinfo{pages}{865} (\bibinfo{year}{1964}),
  ISSN \bibinfo{issn}{0022-3697}.

\bibitem[{\citenamefont{Mao et~al.}(1967)\citenamefont{Mao, Bassett, and
  Takahashi}}]{mao-272}
\bibinfo{author}{\bibfnamefont{H.-K.} \bibnamefont{Mao}},
  \bibinfo{author}{\bibfnamefont{W.~A.} \bibnamefont{Bassett}},
  \bibnamefont{and}
  \bibinfo{author}{\bibfnamefont{T.}~\bibnamefont{Takahashi}},
  \bibinfo{journal}{J. Appl. Phys.} \textbf{\bibinfo{volume}{38}},
  \bibinfo{pages}{272} (\bibinfo{year}{1967}).

\bibitem[{\citenamefont{Cort et~al.}(1982)\citenamefont{Cort, Taylor, and
  Willis}}]{Cort82}
\bibinfo{author}{\bibfnamefont{G.}~\bibnamefont{Cort}},
  \bibinfo{author}{\bibfnamefont{R.~D.} \bibnamefont{Taylor}},
  \bibnamefont{and} \bibinfo{author}{\bibfnamefont{J.~O.}
  \bibnamefont{Willis}}, \bibinfo{journal}{J. Appl. Phys.}
  \textbf{\bibinfo{volume}{53}}, \bibinfo{pages}{2064} (\bibinfo{year}{1982}).

\bibitem[{\citenamefont{Giles et~al.}(1971)\citenamefont{Giles, Longenbach, and
  Marder}}]{Giles71}
\bibinfo{author}{\bibfnamefont{P.~M.} \bibnamefont{Giles}},
  \bibinfo{author}{\bibfnamefont{M.~H.} \bibnamefont{Longenbach}},
  \bibnamefont{and} \bibinfo{author}{\bibfnamefont{A.~R.}
  \bibnamefont{Marder}}, \bibinfo{journal}{J. Appl. Phys.}
  \textbf{\bibinfo{volume}{42}}, \bibinfo{pages}{4290} (\bibinfo{year}{1971}).

\bibitem[{\citenamefont{Katanin et~al.}(2010)\citenamefont{Katanin, Poteryaev,
  Efremov, Shorikov, Skornyakov, Korotin, and Anisimov}}]{Katanin10}
\bibinfo{author}{\bibfnamefont{A.~A.} \bibnamefont{Katanin}},
  \bibinfo{author}{\bibfnamefont{A.~I.} \bibnamefont{Poteryaev}},
  \bibinfo{author}{\bibfnamefont{A.~V.} \bibnamefont{Efremov}},
  \bibinfo{author}{\bibfnamefont{A.~O.} \bibnamefont{Shorikov}},
  \bibinfo{author}{\bibfnamefont{S.~L.} \bibnamefont{Skornyakov}},
  \bibinfo{author}{\bibfnamefont{M.~A.} \bibnamefont{Korotin}},
  \bibnamefont{and} \bibinfo{author}{\bibfnamefont{V.~I.}
  \bibnamefont{Anisimov}}, \bibinfo{journal}{Phys. Rev. B}
  \textbf{\bibinfo{volume}{81}}, \bibinfo{pages}{045117}
  (\bibinfo{year}{2010}).

\bibitem[{\citenamefont{Stixrude et~al.}(1994)\citenamefont{Stixrude, Cohen,
  and Singh}}]{Stixrude94}
\bibinfo{author}{\bibfnamefont{L.}~\bibnamefont{Stixrude}},
  \bibinfo{author}{\bibfnamefont{R.~E.} \bibnamefont{Cohen}}, \bibnamefont{and}
  \bibinfo{author}{\bibfnamefont{D.~J.} \bibnamefont{Singh}},
  \bibinfo{journal}{Phys. Rev. B} \textbf{\bibinfo{volume}{50}},
  \bibinfo{pages}{6442} (\bibinfo{year}{1994}).

\bibitem[{\citenamefont{Burgers}(1934)}]{burgers1934}
\bibinfo{author}{\bibfnamefont{W.~G.} \bibnamefont{Burgers}},
  \bibinfo{journal}{Physica} \textbf{\bibinfo{volume}{1}}, \bibinfo{pages}{561
  } (\bibinfo{year}{1934}).

\bibitem[{\citenamefont{Tsetseris}(2005)}]{Tsetseris05}
\bibinfo{author}{\bibfnamefont{L.}~\bibnamefont{Tsetseris}},
  \bibinfo{journal}{Phys. Rev. B} \textbf{\bibinfo{volume}{72}},
  \bibinfo{pages}{012411} (\bibinfo{year}{2005}).

\bibitem[{\citenamefont{Tsunoda}(1989)}]{Tsunoda89}
\bibinfo{author}{\bibfnamefont{Y.}~\bibnamefont{Tsunoda}},
  \bibinfo{journal}{J. Phys.: Condens. Matter}
  \textbf{\bibinfo{volume}{1}}, \bibinfo{pages}{10427} (\bibinfo{year}{1989}).

\bibitem[{\citenamefont{Hawreliak et~al.}(2011)\citenamefont{Hawreliak,
  El-Dasher, Lorenzana, Kimminau, Higginbotham, Nagler, Vinko, Murphy,
  Whitcher, Wark et~al.}}]{Hawreliak11}
\bibinfo{author}{\bibfnamefont{J.~A.} \bibnamefont{Hawreliak}},
  \bibinfo{author}{\bibfnamefont{B.}~\bibnamefont{El-Dasher}},
  \bibinfo{author}{\bibfnamefont{H.}~\bibnamefont{Lorenzana}},
  \bibinfo{author}{\bibfnamefont{G.}~\bibnamefont{Kimminau}},
  \bibinfo{author}{\bibfnamefont{A.}~\bibnamefont{Higginbotham}},
  \bibinfo{author}{\bibfnamefont{B.}~\bibnamefont{Nagler}},
  \bibinfo{author}{\bibfnamefont{S.~M.} \bibnamefont{Vinko}},
  \bibinfo{author}{\bibfnamefont{W.~J.} \bibnamefont{Murphy}},
  \bibinfo{author}{\bibfnamefont{T.}~\bibnamefont{Whitcher}},
  \bibinfo{author}{\bibfnamefont{J.~S.} \bibnamefont{Wark}},
  \bibnamefont{et~al.}, \bibinfo{journal}{Phys. Rev. B}
  \textbf{\bibinfo{volume}{83}}, \bibinfo{pages}{144114}
  (\bibinfo{year}{2011}).

\bibitem[{\citenamefont{Sanborn et~al.}(2011)\citenamefont{Sanborn, Ludwig,
  Rogers, and Sutton}}]{Sanborn11}
\bibinfo{author}{\bibfnamefont{C.}~\bibnamefont{Sanborn}},
  \bibinfo{author}{\bibfnamefont{K.~F.} \bibnamefont{Ludwig}},
  \bibinfo{author}{\bibfnamefont{M.~C.} \bibnamefont{Rogers}},
  \bibnamefont{and} \bibinfo{author}{\bibfnamefont{M.}~\bibnamefont{Sutton}},
  \bibinfo{journal}{Phys. Rev. Lett.} \textbf{\bibinfo{volume}{107}},
  \bibinfo{pages}{015702} (\bibinfo{year}{2011}).

\bibitem[{\citenamefont{Bl\"ochl}(1994)}]{PhysRevB.50.17953}
\bibinfo{author}{\bibfnamefont{P.~E.} \bibnamefont{Bl\"ochl}},
  \bibinfo{journal}{Phys. Rev. B} \textbf{\bibinfo{volume}{50}},
  \bibinfo{pages}{17953} (\bibinfo{year}{1994}).

\bibitem[{\citenamefont{Torrent et~al.}(2008)\citenamefont{Torrent, Jollet,
  Bottin, Z\'erah, and Gonze}}]{Torrent2008337}
\bibinfo{author}{\bibfnamefont{M.}~\bibnamefont{Torrent}},
  \bibinfo{author}{\bibfnamefont{F.}~\bibnamefont{Jollet}},
  \bibinfo{author}{\bibfnamefont{F.}~\bibnamefont{Bottin}},
  \bibinfo{author}{\bibfnamefont{G.}~\bibnamefont{Z\'erah}}, \bibnamefont{and}
  \bibinfo{author}{\bibfnamefont{X.}~\bibnamefont{Gonze}},
  \bibinfo{journal}{Comput. Mater. Sci.}
  \textbf{\bibinfo{volume}{42}}, \bibinfo{pages}{337 } (\bibinfo{year}{2008}).

\bibitem[{\citenamefont{Gonze et~al.}(2009)\citenamefont{Gonze, Amadon,
  Anglade, Beuken, Bottin, Boulanger, Bruneval, Caliste, Caracas, Côté
  et~al.}}]{Gonze09}
\bibinfo{author}{\bibfnamefont{X.}~\bibnamefont{Gonze}},
  \bibinfo{author}{\bibfnamefont{B.}~\bibnamefont{Amadon}},
  \bibinfo{author}{\bibfnamefont{P.-M.} \bibnamefont{Anglade}},
  \bibinfo{author}{\bibfnamefont{J.-M.} \bibnamefont{Beuken}},
  \bibinfo{author}{\bibfnamefont{F.}~\bibnamefont{Bottin}},
  \bibinfo{author}{\bibfnamefont{P.}~\bibnamefont{Boulanger}},
  \bibinfo{author}{\bibfnamefont{F.}~\bibnamefont{Bruneval}},
  \bibinfo{author}{\bibfnamefont{D.}~\bibnamefont{Caliste}},
  \bibinfo{author}{\bibfnamefont{R.}~\bibnamefont{Caracas}},
  \bibinfo{author}{\bibfnamefont{M.}~\bibnamefont{Côté}}, \bibnamefont{et~al.},
  \bibinfo{journal}{Comput. Phys. Commun.}
  \textbf{\bibinfo{volume}{180}}, \bibinfo{pages}{2582 }
  (\bibinfo{year}{2009}).

\bibitem[{\citenamefont{Monkhorst and Pack}(1976)}]{PhysRevB.13.5188}
\bibinfo{author}{\bibfnamefont{H.~J.} \bibnamefont{Monkhorst}}
  \bibnamefont{and} \bibinfo{author}{\bibfnamefont{J.~D.} \bibnamefont{Pack}},
  \bibinfo{journal}{Phys. Rev. B} \textbf{\bibinfo{volume}{13}},
  \bibinfo{pages}{5188} (\bibinfo{year}{1976}).

\bibitem[{\citenamefont{Methfessel and Paxton}(1989)}]{PhysRevB.40.3616}
\bibinfo{author}{\bibfnamefont{M.}~\bibnamefont{Methfessel}} \bibnamefont{and}
  \bibinfo{author}{\bibfnamefont{A.~T.} \bibnamefont{Paxton}},
  \bibinfo{journal}{Phys. Rev. B} \textbf{\bibinfo{volume}{40}},
  \bibinfo{pages}{3616} (\bibinfo{year}{1989}).

\bibitem[{Note2()}]{Note2}
\bibinfo{note}{AFMI is the AFM coupling of layers in the 0001 direction,
  whereas AFMII is an AFM coupling inside the 0001 plane: See, e.g., Fig 1 of
  Ref.\ 8}.

\bibitem[{\citenamefont{Steinle-Neumann
  et~al.}(1999)\citenamefont{Steinle-Neumann, Stixrude, and Cohen}}]{Steinle99}
\bibinfo{author}{\bibfnamefont{G.}~\bibnamefont{Steinle-Neumann}},
  \bibinfo{author}{\bibfnamefont{L.}~\bibnamefont{Stixrude}}, \bibnamefont{and}
  \bibinfo{author}{\bibfnamefont{R.~E.} \bibnamefont{Cohen}},
  \bibinfo{journal}{Phys. Rev. B} \textbf{\bibinfo{volume}{60}},
  \bibinfo{pages}{791} (\bibinfo{year}{1999}).

\bibitem[{\citenamefont{Lookman et~al.}(2008)\citenamefont{Lookman, Saxena, and
  Albers}}]{Lookman2008}
\bibinfo{author}{\bibfnamefont{T.}~\bibnamefont{Lookman}},
  \bibinfo{author}{\bibfnamefont{A.}~\bibnamefont{Saxena}}, \bibnamefont{and}
  \bibinfo{author}{\bibfnamefont{R.~C.} \bibnamefont{Albers}},
  \bibinfo{journal}{Phys. Rev. Lett.} \textbf{\bibinfo{volume}{100}},
  \bibinfo{pages}{145504} (\bibinfo{year}{2008}).

\end{thebibliography}

\end{document}